# Ultra-Efficient Conversion of Microwave into Ultrasound Wave through a Split Ring Resonator

Lu Lan[1], Yueming Li[2], Tiffany Yang-Tran[1], Yingchun Cao[3], Ji-Xin Cheng[1,3,4]*

1. Department of Biomedical Engineering, Boston University, Boston, MA 02215, USA

2. Department of Mechanical Engineering, Boston University, Boston, MA 02215, USA

3. Department of Electrical & Computer Engineering, Boston University, Boston, MA 02215, USA

4. Photonics Center, Boston University, Boston, MA 02215 USA

Correspondence to: jxcheng@bu.edu

**Abstract**

Thermo-elastic conversion of electromagnetic wave into ultrasound wave has enabled diverse biomedical applications such as photoacoustic imaging. Microwave, which has ~10 cm long wavelength, can penetrate deeper into tissue than photons, heralding exciting applications such as deep tissue imaging via thermo-acoustic tomography. However, the thermo-acoustic conversion efficiency is very low even with an exogenous contrast agent such as carbon nanotube. Here, we break this low conversion limit through using a split ring resonator (SRR) to effectively collect and concentrate the microwave energy into a sub-millimeter hot spot and subsequently convert the energy into ultrasound wave. Our SRR achieves over three orders of magnitude higher thermo-acoustic conversion efficiency than commonly used thermo-acoustic contrast agents. We further harness the SRR as a wireless, battery-free ultrasound emitter placed under a breast phantom. These results promise exciting potential of SRR for precise thermo-acoustic localization and modulation of subjects in deep tissue.

**Introduction**

The photoacoustic (PA) effect, first reported by Bell in 1880 when he invented the photo-phone [1], describes the generation of sound wave through pulsed light absorption by a material. Close to a century later, Bowen envisioned the use of this phenomenon for imaging by excitation using ionizing radiation (e.g., X-ray) or nonionizing radiation (e.g., radiowave and microwave) [2]. With advances in laser technology, instrumentation, and algorithm, PA imaging has now become a multiscale imaging tool from microscopic to macroscopic domains[3-10]. Apart from numerous imaging applications, the PA effect has been recently utilized as a versatile ultrasonic source for ultrasound imaging [11], tissue localization [12], ablation [13], and neuro-modulation [14]. Nevertheless, the dissipation limit of photons in a tissue fundamentally prevents PA applications over 7-cm depth, at which photons are diminished to none by the strong tissue scattering[15,16]. Even with near-infrared light that has less scattering effect and high energy fluence of ~60 mJ/cm$^2$ [17], PA imaging barely reaches 5-7 cm depth. Active PA modulation at such depth is even more challenging to achieve.

Compared to light of sub-micrometer wavelength, microwave at GHz frequency has centimeter long wavelength and thus much weaker tissue scattering, allowing deep penetration for tissue imaging and modulation. Deep tissue and transcranial imaging have been achieved with microwave tomography [18] and thermo-acoustic (TA) imaging [19]. However, the absorption of microwave by tissue is very low compared to the water absorption background, giving a dim image contrast [18,20]. Numerous endeavors have been made to develop exogenous contrast agents with enhanced microwave absorption[20-23] and acoustic conversion efficiency [24]. However, only a maximum of 1-2 times improvement was achieved [23]. Given the low absorption by either endogenous or exogenous agents, microwave for deep tissue applications, such as tumor localization and tissue modulation, is limited.

Here, we report a resonance antenna approach that concentrates microwave energy into a sub-millimeter volume for highly efficient thermo-acoustic generation. Through converting the energy of free propagating radiation into localized energy or vice versa, antennas have been widely used in radiowave and microwave applications and later extended in optics domain, with applications covering photodetection [25,26], light emission [27], photovoltaic [28] and spectroscopy [29,30]. By leveraging the strong local field confinement enabled by a split ring resonator (SRR) --- a building block of microwave metamaterials [31-34], we demonstrate conversion of microwave into ultrasound at an unprecedented conversion efficiency that is three orders of magnitude higher than reported TA contrast agents [20,23]. Although the minimum utilized peak power of 100 W is three orders lower than that used for TA imaging [35], our SRR generates a strong ultrasound close to 40 dB signal-to-noise ratio without averaging. The reported microwave-resonant ultrasound emitter promises broad applications in deep tissue localization and modulation in a wireless and battery-free manner.

## Results

**Highly efficient conversion of microwave into ultrasound via a split ring resonator.**

Similar to the photoacoustic effect, the thermo-acoustic signal generated by an electromagnetic (EM) absorber is proportional to the absorbed energy and its Grüneisen parameter, assuming the thermal and stress confinement are satisfied [36]. The electromagnetic energy $E_{ab}$ absorbed by a volume of tissue depends on the electromagnetic properties of tissue, and is described by Poynting's relation of energy conservation [37,38], as shown below.

$$E_{ab} = \int_V \omega\mu_0\mu_r'' H \cdot H^* dV + \int_V \omega\varepsilon_0\varepsilon_r'' E \cdot E^* dV + \int_V \sigma_c E \cdot E^* dV \tag{1}$$

Where the volume $V$ of tissue has complex permittivity $\varepsilon = \varepsilon_0(\varepsilon_r' - j\varepsilon_r'')$, complex permeability $\mu = \mu_0(\mu_r' - j\mu_r'')$, and ionic conductivity $\sigma_c$. The subscript 0 marks the values of parameters of vacuum, and the subscript $r$ stands for the values of parameters relative to that of vacuum. $E$ is the electric field (V/m), $H$ is the magnetic field (A/m), and $\omega$ is the frequency of the EM wave. These three terms on the right side of **equation (1)** are the power absorbed due to magnetic loss, dielectric polarization loss and joule heating, respectively. It is clearly seen that the power absorbed will be significantly boosted if the local $E$ and $H$ can be enhanced.

Therefore, we attempted to use a microwave resonant antenna to generate a local hotspot of electric field to boost the acoustic generation with a short microwave excitation pulse via the thermo-acoustic effect. We first experimentally examined ultrasound generation from a graphite rod of different lengths placed in oil, given a 1.0 μs microwave excitation pulse of only 100 W peak power at 2.2 GHz (**Fig. S1 a,b**). Surprisingly, the signal is dramatically enhanced when the rod is at a certain length ~39 mm, with an intensity peak at 2.2 GHz corresponding to a microwave wavelength of ~ 78 mm in oil (**Fig. S1 c,d**). This observation suggests that such graphite rod acts as a λ/2 dipole antenna which concentrates the $E$ field at its tips, showing that resonance can greatly improve the conversion efficiency of microwave into ultrasound wave.

To realize highly efficient conversion of microwave into ultrasound at sub-millimeter spatial precision, we chose a single SRR---a building unit for microwave metamaterial. The single split ring is a metallic ring with a small split placed in a medium (**Fig. 1a**). The inset is a photo of the ring used. The SRR can be generally modeled as a *LC* resonance circuit: an inductor *L* formed by the metallic ring and a capacitor *C* formed by the split gap as well as some surface capacitance [39]. When the SRR resonates with microwave excitation, strong electric field is confined inside the capacitor, i.e. the ring gap, and a hotspot subsequently forms if there is microwave absorption by nearby medium. Once given a short microwave excitation pulse of nanosecond to microsecond duration, the localized hotspot causes a transient local volumetric expansion, resulting in an ultrasound wave generation.

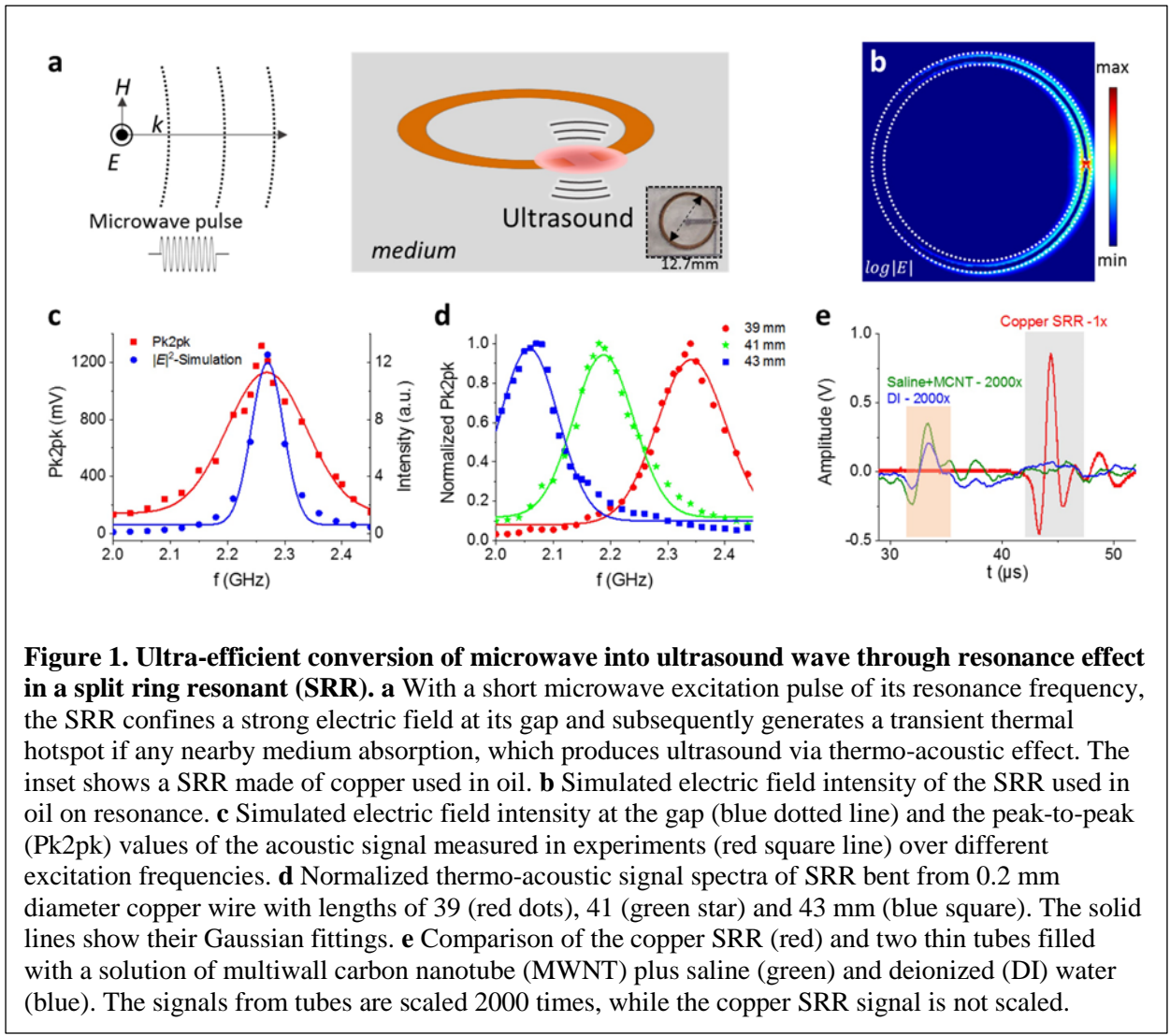

**Figure 1. Ultra-efficient conversion of microwave into ultrasound wave through resonance effect in a split ring resonant (SRR). a** With a short microwave excitation pulse of its resonance frequency, the SRR confines a strong electric field at its gap and subsequently generates a transient thermal hotspot if any nearby medium absorption, which produces ultrasound via thermo-acoustic effect. The inset shows a SRR made of copper used in oil. **b** Simulated electric field intensity of the SRR used in oil on resonance. **c** Simulated electric field intensity at the gap (blue dotted line) and the peak-to-peak (Pk2pk) values of the acoustic signal measured in experiments (red square line) over different excitation frequencies. **d** Normalized thermo-acoustic signal spectra of SRR bent from 0.2 mm diameter copper wire with lengths of 39 (red dots), 41 (green star) and 43 mm (blue square). The solid lines show their Gaussian fittings. **e** Comparison of the copper SRR (red) and two thin tubes filled with a solution of multiwall carbon nanotube (MWNT) plus saline (green) and deionized (DI) water (blue). The signals from tubes are scaled 2000 times, while the copper SRR signal is not scaled.

The first SRR tested is a machined copper ring placed in oil. It has a diameter of 12.7 mm, wire width of 0.8 mm, gap of 0.4 mm and thickness of 0.2 mm. The oil is chosen as the medium for it has small but none-zero microwave absorption and high thermal expansion coefficient. Through COMSOL Multi-physics simulation with no absorption considered, the SRR was found to resonate at 2.27 GHz. The simulated intensity map of the *E* field in log scale on the SRR at resonance clearly shows a local hotspot in the ring gap (**Fig. 1b**). When comparing the measured peak-to-peak (Pk2pk) values of the normalized acoustic signal spectra to the simulated *E* field intensity from the gap (**Fig. 1c**), it is seen that the resonance peak from the acoustic measurement matches the simulation result, while the acoustic measurement shows a broader full-width at half-maximum (FWHM) of 0.16 GHz than 0.07 GHz from simulation. It is understandable that the absorption effect which generates heat and acoustic wave broadens the resonance peak, as others reported [40]. To further confirm the resonance effect, we bent a copper wire of 0.2 mm diameter with different lengths of 39, 41 and 43 mm into a split ring with comparable gap sizes. With longer lengths of copper wire bent into a ring, the inductance of the SRR increases and results in a lower resonance frequency. It is consistent with our experimental observations that the resonance frequency red shifts with longer ring perimeters (**Fig. 1d**).

Next, we compared the ultrasound signal generated by the copper SRR to that by de-ionized (DI) water or multi-wall carbon nanotubes (MWCNT) mixed with saline solution in a thin polyurethane tube. The tube length was designed to be 39 mm to match the circumference of the SRR. Acoustic signals from tubes and SRR were recorded under the same experimental conditions. Due to the low and noisy signals from the tubes with a low excitation peak-power of 100 W, the signals were averaged for 8192 times and smoothed in post-processing to obtain good signal to noise ratio (SNR), whereas the signal from SRR was only averaged by 2 times and had an SNR of ~160. For a clear head-to-head comparison, we scaled the signal from both tubes by 2000 times and plotted them with the unscaled SRR signal at its resonance frequency (**Fig. 1e**). It is observed that the copper SRR generated an acoustic signal that is more than three orders of magnitude higher than that from MWCNT, a commonly used TA imaging contrast agent [20,23]. These data collectively show that the resonance in SRR tremendously boosts up the conversion efficiency of microwave into ultrasound wave.

**The microwave-thermal resonance in SRR is confirmed by thermal imaging**

To confirm the thermal hotpot generated at the ring gap, we performed thermal imaging on the SRR with microwave excitation. First, we mounted the SRR on a thermal paper (Brother LB3635, USA) that permanently darkens around 85 ℃ to qualitatively visualize the thermal hotpot when the SRR was embedded in a bulky oil medium with microwave excitation (**Fig. 2a,b**). The microwave source was running in continuous mode and turned on for approximately 250 ms. Dark spots were only formed at the ring gap (**Fig. 2c**) with the largest dark spot formed with 2.27 GHz frequency. This is consistent with the resonance frequency measured by the acoustic detection in **Fig. 1c**. This result qualitatively confirms the microwave-thermal resonance and the hotspot generation in the ring gap of SRR.

To quantitatively visualize the temperature change in the SRR when microwave is illuminated, we mounted the SRR on a thin plastic film andflipped the film so that the ring was floating on the surface of a small oil-filled container. By doing so, the ring was shallowly immersed in oil and the mid-infrared light radiated from the SRR can be captured by a thermal camera. Note that this oil-air interface shifts the resonance of SRR to higher frequency, 2.49 GHz, compared to that in the case of bulky oil medium, which was verified with our numerical simulation (**Fig. S3**). The microwave source was running in continuous mode and turned on for approximately 250 ms. The thermal camera captures the heating and relaxing process with a frame rate of 30 Hz for 20 seconds. The experiments were done with different microwave frequencies: 2.00 (off-resonance), 2.35 (near-resonance), and 2.49 GHz (on-resonance).

Before the microwave heating, no hotspot or contrast showed up at time 0.0 s (**Fig. 2d-f**). After the excitation was turned on for 250 ms, a hotpot was observed in the ring gap when the microwave frequency matches the resonance frequency of SRR (**Fig. 2g**); while very weak contrast showed up at the gap with 2.0 GHz microwave excitation (**Fig. 2h,i**). The dynamic heating process can be viewed in **Movie S1-3**. **Fig. 2j** shows the temporal temperature profile inside the gap in all excitation frequencies. It shows that a temperature increase of 69.4 K was observed for 250 ms heating time at the ring gap when it was on resonance, while the temperature rise was less than 0.3 K when the ring was off-resonance. Notably, if assuming a linear heating process inside the gap, the temperature rise is estimated to be less than 0.3 mK given a 1.0 µs microwave excitation to generate ultrasound through the thermo-acoustic effect.

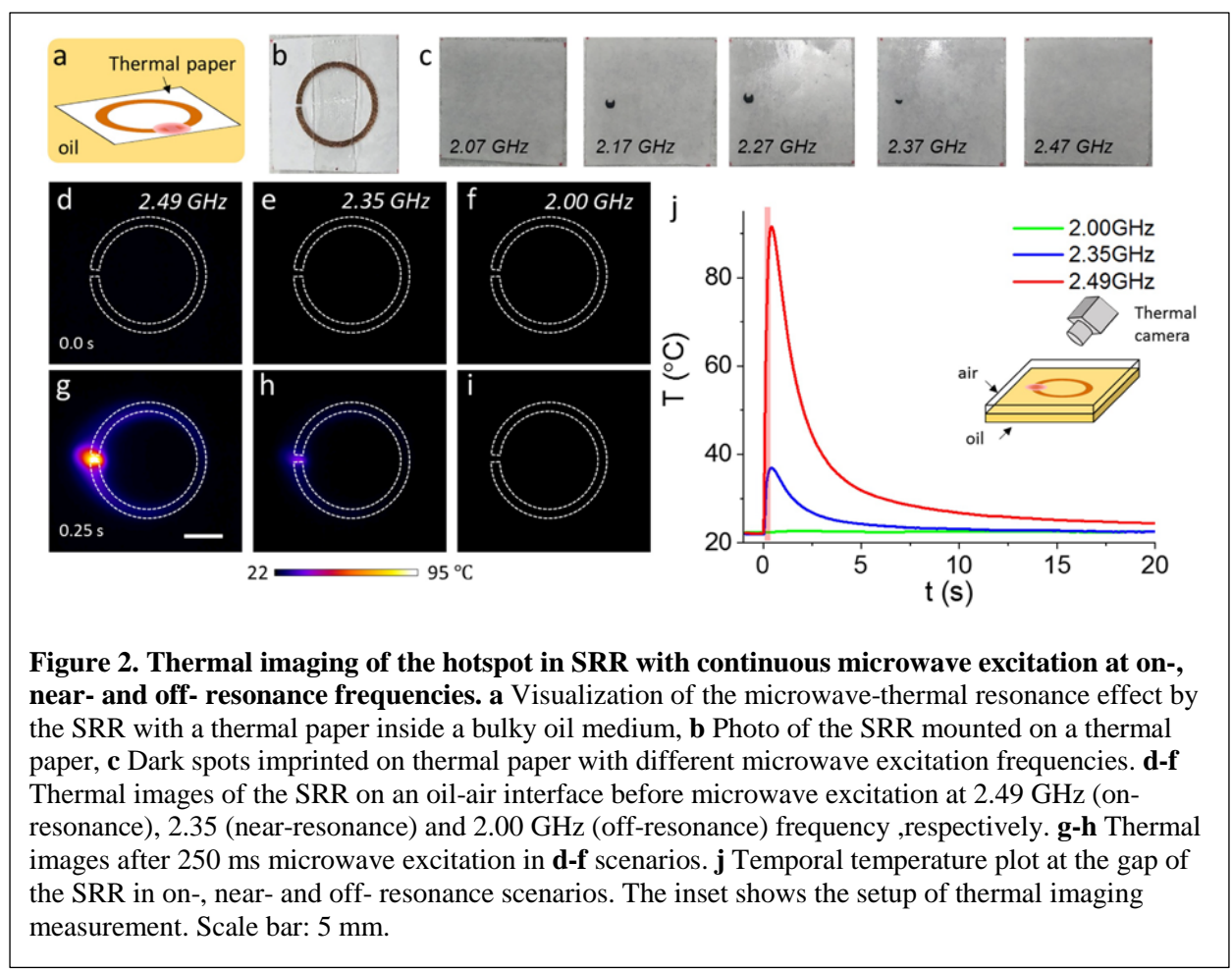

**Figure 2. Thermal imaging of the hotspot in SRR with continuous microwave excitation at on-, near- and off- resonance frequencies. a** Visualization of the microwave-thermal resonance effect by the SRR with a thermal paper inside a bulky oil medium, **b** Photo of the SRR mounted on a thermal paper, **c** Dark spots imprinted on thermal paper with different microwave excitation frequencies. **d-f** Thermal images of the SRR on an oil-air interface before microwave excitation at 2.49 GHz (on-resonance), 2.35 (near-resonance) and 2.00 GHz (off-resonance) frequency ,respectively. **g-h** Thermal images after 250 ms microwave excitation in **d-f** scenarios. **j** Temporal temperature plot at the gap of the SRR in on-, near- and off- resonance scenarios. The inset shows the setup of thermal imaging measurement. Scale bar: 5 mm.

**Visualization of acoustic wave generation from the SRR**

We constructed a thermo-acoustic (TA) imaging system to visually verify the ultrasound generation from the ring gap. A TA imaging system with a 128-channel transducer array (L7-4, ATL) was built for experiments here. To match the frequency band of the transducer array, which is 4 to 7 MHz, we used a 0.1 µs microwave excitation pulse at 2.33 GHz. By doing so, a cross-sectional TA image (parallel to the x-y plane) is acquired. We then translated the transducer array to obtain z stack images to obtain 3D profile of the acoustic emission from SRR. The SRR was placed in the oil medium in two configurations with the magnetic field $H$ always perpendicular to the ring plane: the first to measure the acoustic emission above the ring plane (**Fig. 3a-g**) and the second to measure the aforementioned in the ring plane (**Fig. 3h-j**). The delay between the excitation pulse and ultrasound detection was adjusted to capture the dynamics of the acoustic field propagating out from the ring gap.

**Fig. 3a** shows the ultrasound (US) image of the SRR in the x-z plane, which was mounted in oil on a thin narrow plastic strip such that it bisected the ring. The corresponding TA image at the x-y plane at time delay 0 was shown in **Fig. 3b**. When merged with the US image (**Fig. 3c**), the origin of acoustic generation was confirmed to be the ring gap. **Fig. 3d** shows the cross sectional view of TA image right above the gap of the SRR. When plotting the line profiles across the center of the gap, FWHM of 0.80 and 0.49 mm were measured in lateral and axial directions, respectively. These dimensions are close to the size of the gap (lateral: 0.4 mm and axial: 0.3 mm), considering the low spatial resolution by the low frequency transducer array used. By tuning the delay of the ultrasound detection relative to the microwave excitation pulse, the acoustic propagation process was visualized. **Fig. 3e-g** shows the TA signal in 3D at time delay of 0, 4 and 8 µs, respectively. **Movies S4 and S5** show the dynamic process of acoustic generation from the ring gap. Next, the acoustic emission in the ring plane was measured by re-orienting the ring as shown in the bottom left of **Fig. 3**. By merging the US (**Fig. 3h**) and TA (**Fig. 3i**) images, the TA signal was also found from the gap (**Fig. 3j**), proving acoustic emission

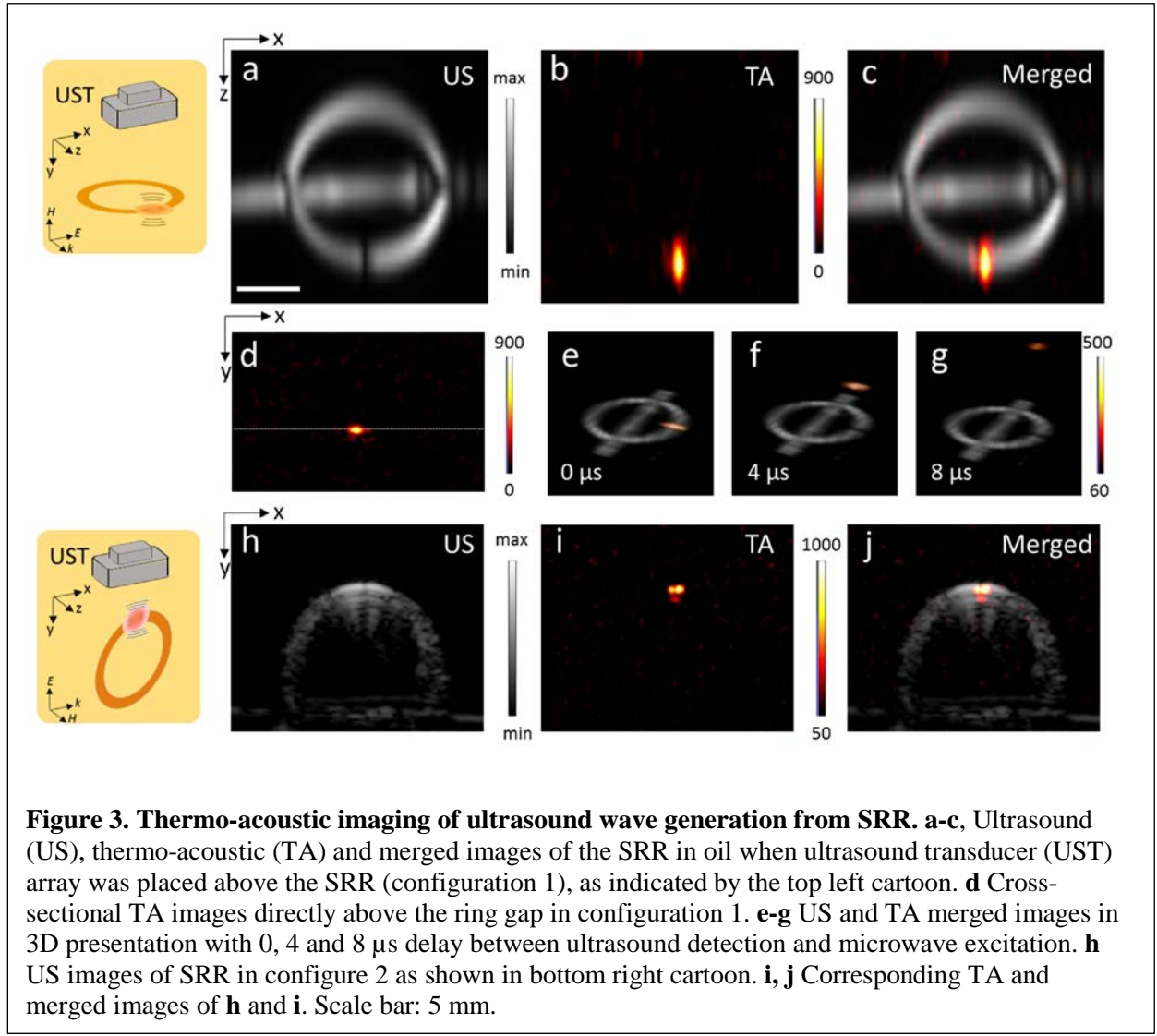

**Figure 3. Thermo-acoustic imaging of ultrasound wave generation from SRR. a-c**, Ultrasound (US), thermo-acoustic (TA) and merged images of the SRR in oil when ultrasound transducer (UST) array was placed above the SRR (configuration 1), as indicated by the top left cartoon. **d** Cross-sectional TA images directly above the ring gap in configuration 1. **e-g** US and TA merged images in 3D presentation with 0, 4 and 8 µs delay between ultrasound detection and microwave excitation. **h** US images of SRR in configure 2 as shown in bottom right cartoon. **i, j** Corresponding TA and merged images of **h** and **i**. Scale bar: 5 mm.

exists in the ring plane. Additionally, **Fig. 3i** shows two small hotspots in the TA image, which corresponds to two flat ends of the gap and agrees with the simulation in **Fig. 1b**. We also oriented the ultrasound probe at oblique angle to the ring plane and obtained TA signals from 0-60 degrees except for 20 degrees (**Fig. S4**). Collectively, this data provides direct evidence of acoustic wave generation from the two adjacent hotspots in the SRR gap following pulsed

**Effect of microwave pulse duration on the thermo-acoustic signal generation**

The excitation pulse duration affects the acoustic signal generation on its amplitude, efficiency, etc. Given the challenges of having a high-energy nanosecond laser source with tunable pulse duration, the effect of pulse duration on acoustic signal generation is not fully investigated [41]. Since changing the pulse duration of microwave excitation can be done electronically, we measured the acoustic signal generated from the SRR with different pulse durations by two single element transducers with 5 and 0.5 MHz center frequencies under the same experimental conditions. **Fig. 4a, c** shows some representative acoustic waveforms over different pulse durations measured by single element transducers of 5 and 0.5 MHz center frequency, respectively. It is seen that the acoustic signal changes with different pulse durations. Specifically, two bipolar acoustic signals appeared with long pulse durations, such as 2 µs excitation pulse in **Fig. 4a** and 5 µs pulse in **Fig. 4c**. Two acoustic pulses generated from the rising and falling edges of the excitation pulse and become discernible when the excitation pulse duration exceeds the response time of the detectors. We plotted the peak-to-peak values of the acoustic signal over the excitation pulse duration in **Fig. 4b, d**. It is observed that the pk2pk values first increase with longer pulse durations since more pulse energy was input on the SRR. Then, the pk2pk value started to drop after a pulse duration of 0.6 µs for the 5 MHz transducer and 1 µs for the 0.5 MHz transducer. It agrees well that the 0.5 MHz detector has a smaller bandwidth than that of the 5 MHz detector and thus needs longer response time to discern the two acoustic signals generated from rising and falling edges of the long excitation pulses. Therefore, excitation pulse duration and detector bandwidth shall be carefully chosen dependent on applications

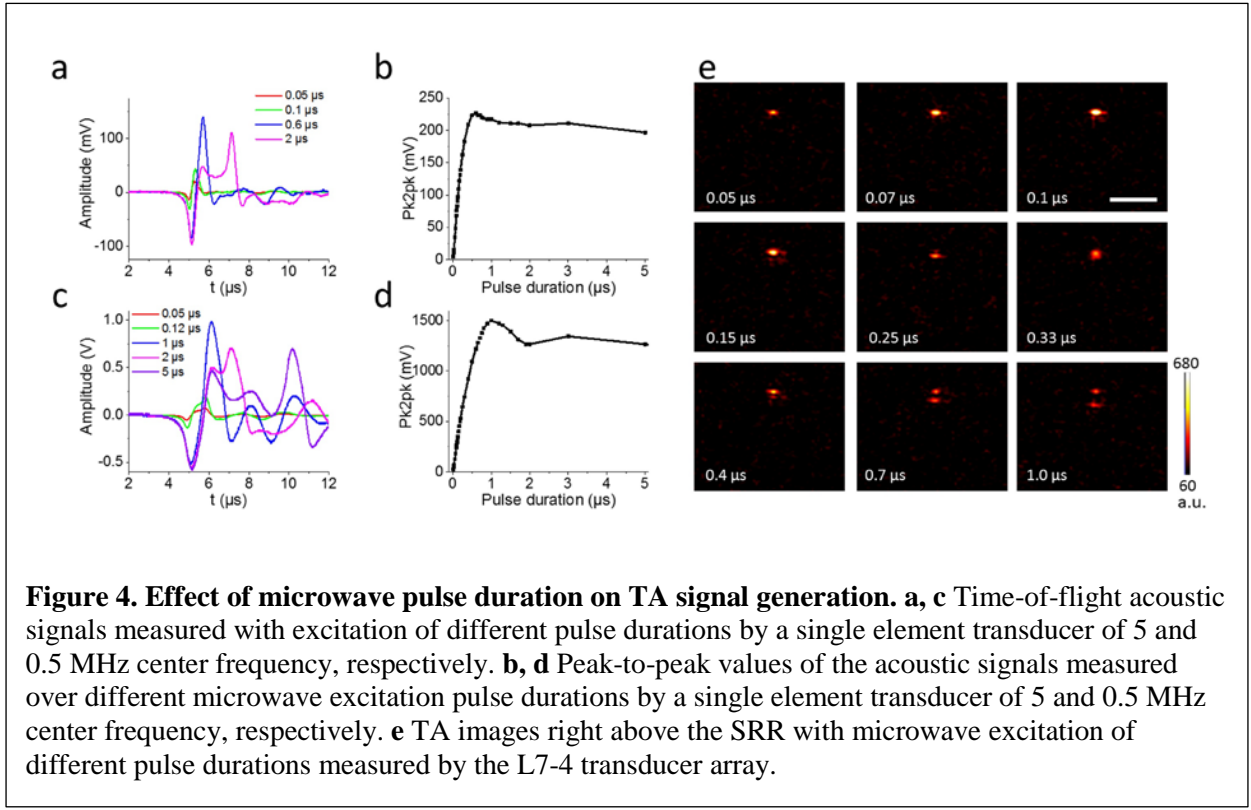

**Figure 4. Effect of microwave pulse duration on TA signal generation. a, c** Time-of-flight acoustic signals measured with excitation of different pulse durations by a single element transducer of 5 and 0.5 MHz center frequency, respectively. **b, d** Peak-to-peak values of the acoustic signals measured over different microwave excitation pulse durations by a single element transducer of 5 and 0.5 MHz center frequency, respectively. **e** TA images right above the SRR with microwave excitation of different pulse durations measured by the L7-4 transducer array.

We further used our TA system to directly visualize this splitting of acoustic emission from the SRR with certain long excitation pulse durations. **Fig. 4e** shows the cross-sectional view of TA images above the gap with different excitation pulse durations. As the pulse duration increased from 0.05 to 0.1 µs, the TA signal intensity reached its maximum (**Fig. S5**). Further increase of pulse duration did not help increase the TA intensity. Interestingly, when the pulse duration exceeded 0.33 µs, the TA spot started to split into two. The splitting effect of the acoustic wave packets can further be obtained with the radio-frequency data retrieved from the ultrasound system module (**Fig. S5**). We note that with the transducer array, the splitting is observed at a shorter microwave pulse duration compared to the single element transducer measurement. This discrepancy is due to the faster temporal response of the transducer array used here, which has a bandwidth of approximately 3.0 MHz. The two TA spots correspond well with the rising and falling edges of the microwave excitation pulse, and the separation between the two split TA spots increased as longer pulse duration was applied. For instance, the peak distance between two TA spots was 0.94, 1.54 and 4.40 mm for excitation pulse duration of 0.7, 1.0 and 3.0 µs, respectively (**Fig. S6**). Through linear fitting, the ultrasound speed was estimated to be 1472 m/s, close to the sound speed of 1480 m/s used in the imaging reconstruction algorithm.

If defining the excitation efficiency to be the peak-to-peak values of the acoustic signal over the excitation pulse duration, we obtained an efficiency curve using 5 and 0.5 MHz transducer (**Fig. S7**). It is found that the excitation efficiency firstly increased to its maximum at certain pulse duration and subsequently drops when the pulse duration increases from 0.05 to 5 µs. The pulse duration for maximum efficiency was found to be around 0.1 µs using the 5 and 0.5 MHz transducer. Collectively, this data shows that the TA signal can be further enhanced with microwave excitation pulse of higher peak power and shorter pulse duration.

**SRR as a wireless and battery-free ultrasound emitter**

As an initial proof-of-concept of a wireless, battery-free, ultrasound emitter, we packaged the SRR in a small thin plastic bag filled with a small volume of canola oil (**Fig. 5a**). The SRR was placed under a breast biopsy ultrasound training phantom. Using a microwave excitation of 1 µs pulse at 2.33 GHz with 1 k Hz repetition rate, we successfully received the ultrasound signal by placing the transducer above the phantom. **Fig. 5b** shows the detected ultrasound signal with 64 times averaging, and the SNR was measured to be about 30. The distance of the ring antenna to the transducer was estimated to be 53 mm, which is close to the thickness of the phantom. When the waveguide was moved away from the breast phantom, a decrease of the peak-to-peak values of the acoustic signal was observed (**Fig. 5c**), but still an acoustic signal of a fair intensity can be obtained at a distance of 150 mm. Next, we placed a card box and a box full of gloves between the waveguide and breast phantom, the signal intensity did not decrease as expected, because the card box and glove box are transparent to microwave (**Fig. 5d, e**). Moreover, the acoustic signal slightly increased when the glove box was inserted. It is also understandable that the glove box has rubber gloves which have a dielectric constant to the silicone breast phantom and acted as a matching layer to couple more microwave energy into the phantom. Thus, placing the SRR in a small volume of oil, it can act as a wireless, battery-free, ultrasound emitter with pulsed microwave excitation of only 10 mW average power.

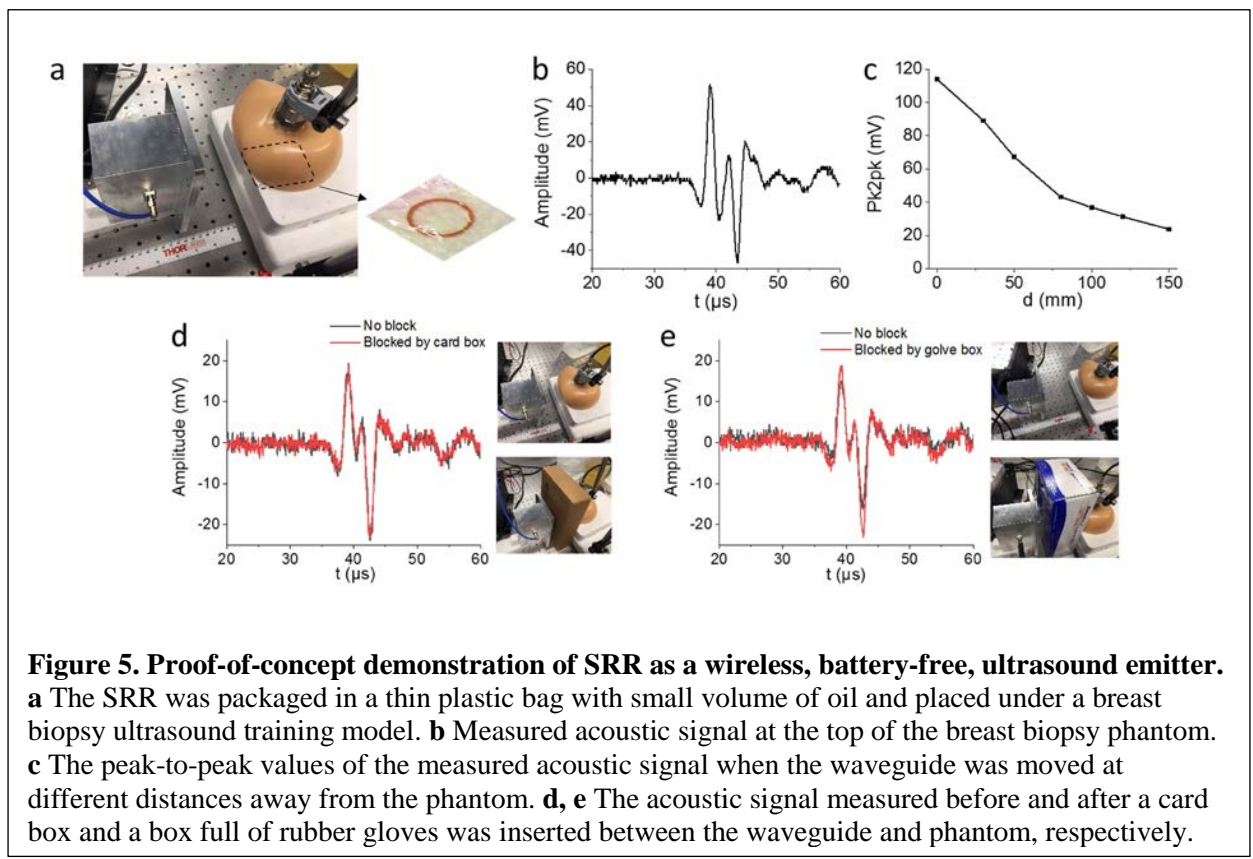

**Figure 5. Proof-of-concept demonstration of SRR as a wireless, battery-free, ultrasound emitter.**
**a** The SRR was packaged in a thin plastic bag with small volume of oil and placed under a breast biopsy ultrasound training model. **b** Measured acoustic signal at the top of the breast biopsy phantom. **c** The peak-to-peak values of the measured acoustic signal when the waveguide was moved at different distances away from the phantom. **d, e** The acoustic signal measured before and after a card box and a box full of rubber gloves was inserted between the waveguide and phantom, respectively.

**Discussion**

In this work, we harnessed an SRR to efficiently collect and convert the microwave energy into heat and then ultrasound wave for potential biomedical applications in deep tissue. Compared to the commonly used TA imaging contrast agent, such as multiwall carbon nanotube, the SRR achieved over three orders higher conversion efficiency in generating ultrasound wave. With an energy fluence as low as 1.7 µJ/cm², the SRR generated strong ultrasonic signal at MHz frequency of over 100 SNR without averaging. Similar resonance mechanisms can be also transferred to optical resonators, such as micro-ring and whispering gallery mode resonator [42,43], to improve the conversion efficiency of light into ultrasound.

Using thermal and TA imaging, we experimentally confirmed that the strongly enhanced electrical field was confined at the submillimeter size gap of the SRR. It shows the capability of SRR to confine energy into submillimeter scale, which is at the scale of $\lambda_0/100$, to achieve active bio-modulation with high spatial resolution. In thermal imaging experiments, over 69 degrees temperature rise was observed when the ring was shallowly submerged in the oil and excited with on-resonance microwave for 250 ms, while it was only 0.3 degrees when tuned to off-resonance excitation. Such efficient submillimeter thermal hotpot generated from the SRR can enable various deep tissue thermal modulation devices and applications with high precision, such as wireless thermal neuron modulation using very short pulse excitation, in which only a local temperature rise over 5 degrees is required [44].

The current work presents the first use of SRR for resonance to enhance the collection and conversion of microwave into ultrasound wave for potential active bio-modulation use. Microwave resonator of other geometries [45,46] or periodic arrays [47] can be tailored for detailed applications. We showed an example of using the SRR as a wireless, battery-free, ultrasound emitter under a breast biopsy training phantom. The SRR can also be applied externally to be a wearable, battery-free, ultrasound emitter for various wearable ultrasound applications, such as wireless ultrasound modulation for pain management [48,49]. Additionally, it provides the foundation to develop a wireless thermo-acoustic guide for lesion localization in soft tissue for precise removal, such as breast conserving surgery.

## Methods

**Numerical simulation of the resonance frequency of SRR.** The simulations were performed using COMSOL Multi-physics 5.3a. In all simulations, the dielectric constant of canola oil was set to 2.4, unless otherwise specified. The excitation wave was provided using a port with plane wave input that has $E$ polarized in $y$ direction and $H$ polarized in $z$ direction. The magnetic field is polarized perpendicular to the SRR plane ($x$-$y$ plane). Scattering conditions were used at the boundaries of the simulated area.

**Measurement of the acoustic signal generated from the SRR.** We placed the machined copper ring into a plastic tank filled with canola oil. A microwave signal generator (9 kHz – 3 GHz, SMB100A, Rohde & Schwarz) was used as the seed microwave generation, and a solid state power amplifier (ZHL-100W-242+, Mini Circuits) was connected to amplify the microwave to 100 Watt peak power. Next, the amplified microwave was delivered through a waveguide (WR430, Pasternack) to the oil tank. The distance from the SRR to the waveguide was about 2 cm. For acoustic generation, the microwave source was operating in pulsed mode, and the pulse duration was 1 µs at 1 kHz repetition rate, if not otherwise specified. A single element ultrasound transducer (SV301, Olympus) with 0.5 MHz center frequency was used to detect the generated acoustic signal. The distance between the transducer and the SRR was approximately 52 mm. The received acoustic signal was first amplified by a pulser/receiver (5072PR, Olympus) at receiving mode with 59 dB gain and 0-10 MHz filter applied. Lastly, the detected ultrasonic signal was read out using an oscilloscope (DS4024, Rigol). To confirm the resonance effect, the frequency of the microwave excitation was scanned from 2-2.5 GHz.

**Comparison of the acoustic signal generated from the SRR and multi-wall carbon nanotubes, a thermo-acoustic contrast agent.** De-ionized (DI) water or multi-wall carbon nanotubes (MWCNT) mixed with saline solution in a thin polyurethane tube, following the protocol in *ref. 20*. The polyurethane tube length was designed to be 39 mm to match with the perimeter of the SRR. Acoustic signals from tubes and SRR were individually recorded by the same 0.5 MHz transducer placed at 52 mm distance away as described above. Due to the low and noisy signals from the tubes with low excitation peak-power of 100W, their signals were averaged 8192 times and smoothed in post-processing to obtain signals of good SNR, while the signal from SRR was only averaged 2 times.

**Thermal imaging.** To visualize the temperature change of the ring when microwave is illuminated, we mounted the ring on a thin plastic film, inverted the film and made it float on a small oil-filled container. By doing so, the ring was shallowly immersed in oil and the mid-infrared light radiated from the ring could be captured by the thermal camera. A thermal camera (A325sc, FLIR) was mounted above and look down at the ring floating in the oil container. The microwave source was running in continuous mode and turned on for 250 ms for thermal imaging experiments. The thermal camera captures the heating and relaxation process with a frame rate of 30 Hz for 20 seconds. A broad band power amplifier (0.7-2.7 GHz, ZHL-100W-272+, Mini Circuits) was later used in the thermal imaging experiments here.

**Thermo-acoustic imaging.** We built a thermo-acoustic imaging system by replacing the single element transducer with a transducer array (L7-4, ATL) and 128-channel ultrasound data acquisition system (Vantage 128, Verasonics). A function generator and a delay generator (9254,

Quantum Composer) worked together to synchronize the microwave excitation and ultrasound detection modules. The function generator outputs a pulse at 20 Hz repetition rate to the microwave signal generator. The delay generator receives the master trigger and adds a controllable delay $t_d$ to trigger the ultrasound data acquisition. The delay can be tuned so that the acoustic field at varying times after ultrasound wave is generated at the ring gap. Note that, it is equivalent to using thebeam forming technique to post-process the raw ultrasound data. The TA images were averaged 200 or 800 times to obtain images of good SNR.

**Effect of pulse duration on TA signal generation measured by single element transducers.**

Microwave excitation at 2.33 GHz with pulse durations from 0.02 to 5 µs were used, and the transducers were placed about 52 mm above the gap of the split ring. Same detection parameters were applied to both cases as above, except that the averaging time was 256 to get a good SNR for short excitation pulses.

**Demonstration of a wireless, battery-free, ultrasound emitter.** As a proof-of-concept demonstration, we put the ring in a small oil bag and placed them under a breast biopsy phantom (BPB170, CAE Healthcare) to act as wireless, battery-free, MHz ultrasound emitter. The same single element transducer (SV301, Olympus) was placed on the top of the breast phantom to detect the generated acoustic signal. Also, the microwave waveguide was moved away from the breast phantom at different distances to explore how distances affected the signal. Additionally, a card box and a box full of gloves were placed between the waveguide and breast phantom to demonstrate a wireless ultrasound emitter.

**Data availability.** The authors declare that all of the data supporting the findings of this study are available within the paper and the supplementary information.

Acknowledgements

The authors thank Dr. Thomas Bifano for his kindness in lending us the thermal imaging camera. We also want to express our gratitude to Dr. Pu Wang for his constructive discussion in the preliminary experimental design. The project is supported by an Ignition Award from Boston University to J.X.C.

Conflict of Interests

The authors declare no conflict of interests.

Author Contributions

L. L. and J.X.C. conceived the idea; L. L. designed the system and carried out the experiments with Y. L. and T.Y.T; Y.C. and T.Y.T conducted the simulations; L. L., Y. L. and T.Y.T analyzed the data; L. L. wrote the manuscript supervised by J.X.C; All authors discussed the results and contributed to the final manuscript.

**Figure and captions**

**Figure 1. Ultra-efficient conversion of microwave into ultrasound wave through resonance effect in a split ring resonant (SRR). a** With a short microwave excitation pulse of its resonance frequency, the SRR confines a strong electric field at its gap and subsequently generates a transient thermal hotspot if any nearby medium absorption, which produces ultrasound via thermo-acoustic effect. The inset shows a SRR made of copper used in oil. **b** Simulated electric field intensity of the SRR used in oil on resonance. **c** Simulated electric field intensity at the gap (blue dotted line) and the peak-to-peak (Pk2pk) values of the acoustic signal measured in experiments (red square line) over different excitation frequencies. **d** Normalized thermo-acoustic signal spectra of SRR bent from 0.2 mm diameter copper wire with lengths of 39 (red dots), 41 (green star) and 43 mm (blue square). The solid lines show their Gaussian fittings. **e** Comparison of the copper SRR (red) and two thin tubes filled with a solution of multiwall carbon nanotube (MWNT) plus saline (green) and deionized (DI) water (blue). The signals from tubes are scaled 2000 times, while the copper SRR signal is not scaled.

**Figure 2. Thermal imaging of the hotspot in SRR with continuous microwave excitation at on-, near- and off- resonance frequencies. a** Visualization of the microwave-thermal resonance effect by the SRR with a thermal paper inside a bulky oil medium, **b** Photo of the SRR mounted on a thermal paper, **c** Dark spots imprinted on thermal paper with different microwave excitation frequencies. **d-f** Thermal images of the SRR on an oil-air interface before microwave excitation at 2.49 GHz (on-resonance), 2.35 (near-resonance) and 2.00 GHz (off-resonance) frequency ,respectively. **g-h** Thermal images after 250 ms microwave excitation in **d-f** scenarios. **j** Temporal temperature plot at the gap of the SRR in on-, near- and off- resonance scenarios. The inset shows the setup of thermal imaging measurement. Scale bar: 5 mm.

**Figure 3. Thermo-acoustic imaging of ultrasound wave generation from SRR. a-c**, Ultrasound (US), thermo-acoustic (TA) and merged images of the SRR in oil when ultrasound transducer (UST) array was placed above the SRR (configuration 1), as indicated by the top left cartoon. **d** Cross-sectional TA images directly above the ring gap in configuration 1. **e-g** US and TA merged images in 3D presentation with 0, 4 and 8 µs delay between ultrasound detection and microwave excitation. **h** US images of SRR in configure 2 as shown in bottom right cartoon. **i, j** Corresponding TA and merged images of **h** and **i**. Scale bar: 5 mm.

**Figure 4. Effect of microwave pulse duration on TA signal generation. a, c** Time-of-flight acoustic signals measured with excitation of different pulse durations by a single element transducer of 5 and 0.5 MHz center frequency, respectively. **b, d** Peak-to-peak values of the acoustic signals measured over different microwave excitation pulse durations by a single element transducer of 5 and 0.5 MHz center frequency, respectively. **e** TA images right above the SRR with microwave excitation of different pulse durations measured by the L7-4 transducer array.

**Figure 5. Proof-of-concept demonstration of SRR as a wireless, battery-free, ultrasound emitter. a** The SRR was packaged in a thin plastic bag with small volume of oil and placed under a breast biopsy ultrasound training model. **b** Measured acoustic signal at the top of the breast biopsy phantom. **c** The peak-to-peak values of the measured acoustic signal when the waveguide was moved at different distances away from the phantom. **d, e** The acoustic signal measured before and after a card box and a box full of rubber gloves was inserted between the waveguide and phantom, respectively.